# An Optimum Scheduling Approach for Creating Optimal Priority of Jobs with Business Values in Cloud Computing

Vivek Sharma

*Department Of Computer Science & Engineering, Jain University, Bangalore-560002, India*
vivek_bsa3@yahoo.com

T. R. Gopalakrishnan Nair

*Research and Industry Incubation Center, Dayananda Sagar Institutions, Bangalore-560078, India*
trgnair@gmail.com, www.trgnair.org

*Abstract—* Realizing an optimal task scheduling by taking into account the business importance of jobs has become a matter of interest in pay and use model of Cloud computing. Introduction of an appropriate model for an efficient task scheduling technique could derive benefit to the service providers as well as clients. In this paper, we have addressed two major challenges which has implications on the performance of the Cloud system. One of the major issues is handling technical aspects of distributing the tasks for targeted gains and the second issue is related to the handling of the business priority for concurrently resolving business complexity related to cloud consumers. A coordinated scheduling can be achieved by considering the weightage of both aspects viz. technical requirements and business requirements appropriately. It can be done in such a way that it meets the QoS requirements of technical domain as well as business domain. Along with the technical priority a business Bp is required in creating a resultant priority which could be given to stages of further processing, like task allocation and arbitration schemes. Here we consider a technical priority Tp that is governed by a semi-adaptive scheduling algorithm whereas the resultant priority is derived in which a Business Priority Bp layer encapsulates the Technical Priority Tp to achieve the overall priority of the incoming tasks. It results in a Hybrid priority creation, which is a combination of both technical priority Tp and business priority Bp. By taking into account the business priority of the jobs it is possible to achieve a higher service level satisfaction for the tasks which are submitted with their native technical priority. With this approach the waiting time of the tasks tends to get reduced and it gives a better service level satisfaction.

## I. INTRODUCTION

Cloud computing is emerging as an important field distinguished from conventional collective computing models by its focus on wide-scale resource sharing across the globe. Consumers and computing platforms are no more restricted geographically. In contrast to grid computing, where traditionally a user needed to first share some resources before he or she could be granted access to a larger pool of shared resources, a cloud computing user needs only pay for the computing services. With cloud computing, new internet services can be developed and deployed without capital acquisitions of hardware or large human integration expenses. Cloud Computing integrates and co-ordinates resources as well as users that operate within different control domains with the goal of delivering variety of computing support of various quality of service (QoS). Cloud Computing is a cluster of distributed computers, which provides on-demand resources and services over a network, usually the internet [1]. It collects all the computing resources and manages them automatically through software. In the process of data analysis for improvement, it integrates the history data and the current data to make the composed information more accurate to provide an intelligent service for its clients and enterprises. The clients need not invest on servers, software's, solutions and other resources for computing [2] in this scenario. The cloud also provides on-demand storage strategies for data intensive scientific applications with pay-as-you-go model [22]. As the cloud offers these advantages, the cloud service provider has to ensure information security or the data security to its clients.

Most of the clouds are built on the top of modern data centers [3] and it is expected that more professional cloud installations may come up when better theoretical models of complex integrations are found proven. It incorporates Infrastructure as a Service (IaaS), Platform as a Service (PaaS), and Software as a Service (SaaS) and provides these services like utilities, so the end users are billed based on their utilization of these resources [5]. [4] Presents the comparison between cloud computing and grid computing. Job scheduling is the core value and aim of grid technology, its aim is to use all kind of resources. It can divide a huge task into a lot of independent and no related sub task and then let every node do the jobs. Even any node fails and doesn't return result it doesn't matter; the whole process will not be affected.. Cloud computing will make huge resource pool through grouping all the resources. But the resource provide by cloud is to complete a special task. A user may apply resource from the resource pool to deploy its application. Cloud computing provides a promising advantage for companies and institutions which have to rely on large scale IT operations in a cost effective way. It enables hiring of the IT utilities like infrastructure, software or platform applications. The preliminary cloud based models have paved way to the users to have access towards more computing power and more applications at an attractive price pattern [15]. The cloud technologies can be used by service providers to create hosted services delivered 'on demand' through 'Public Cloud', and by organizations directly to create the aforementioned 'Private Clouds' i.e. flexible pool of resources within their own data centers. Either way the end result is a 'utility' approach for providing IT resources and the application functionality required by the organization to support its business [20]. The base of cloud computing is in the virtualization, distribution and dynamic extendibility, where virtualization is the main feature [12]. Most software and hardware have extended support to virtualization. An IT resource could be software, hardware, operating system and net storage and manage them in the cloud computing platform.

Task scheduling is defined as the process of making scheduling decisions involving resources over multiple administrative domains. In general a job in Cloud computing can be defined as an entity that needs a resource, from a resource request or a` set of applications for processing. A resource which can be scheduled can be a machine instance, data storage device, an application and an environment etc. To schedule a submitted job to the best resource that a job can use typically involves a computing instance, a slice of data storage etc. Many research works have focused on the task-scheduling algorithms for computing demands to achieve the best performance as much as possible together with QoS. Generally a job comes with pre-defined deadlines. The QoS of a job is satisfied if it finishes on or before the specified deadline while the QoS decreases as the excess time over deadline increases. A scheduling strategy should satisfy both the QoS requirement and business requirement. In this paper we have proposed a semi-adaptive scheduling approach along with the business priority consideration. The resources in a cloud are not restricted to hardware but it can also be software services or web services in various forms of instances [7]. [17] put forward two different but related type of clouds and some advantages and disadvantages. One of them is those that provide computing instances on demand and other is that provide computing capacity on demand. Both use similar machines, but the first is designed to scale out by providing additional computing instances, whereas the second is designed to support data-or-compute-intensive applications via scaling capacity. Cloud computing provide several important benefits over today's dominant model in which an enterprise purchases computers a rack at a time and operates them themselves. A cloud computing usage based pricing model offers several advantages, including reduce capital expense, a low barrier to entry, and the ability to scale up as demand requires, as well as to support brief surges in capacity hence the unit cost for cloud based services is often lower than the cost if the services were provided directly by the organization itself. Cloud computing has some disadvantages as well. First, because cloud services are often remote they can suffer the latency and bandwidth related issues associated with any remote application. Second, because hosted cloud services serve multiple costumers, various issues related to multiple costumers sharing the same piece of hardware can arise. For e.g. if one user's application compromises the system, it can also compromise application of other users that share the same system. Also, having data accessible to third parties can present security, compliance , and regulatory issues. Cloud computing is seen as a layered model, i.e. storage cloud provides storage services, a data cloud provides data management services and a compute cloud provide computational services. They are layered to create a stack of cloud services that acts as a computing platform for developing cloud based applications. This is assumed to be largely not a part of the priority formulation manger but a part of the resources management system of cloud which uses the priority model discussed here. This paper is structured as follows, in section II we present the research background, section III presents the framework and the proposed methodology. The performance analysis and results are presented in section IV. Section V presents the conclusion and future work.

## II. RESEARCH BACKGROUND

Task scheduling is one of the significant areas in Cloud Computing. It is similar to Grid Computing. In Cloud Computing scheduling is to be more adaptive catering to the fixed patterns of the Grid. Cloud computing resource ultimately will transform to a model of high fluidity and open source selection rather than fixed grid architecture and heavily bound format of primitive resource scenarios. Task scheduling system is responsible for selecting the best suitable resources for the task in Cloud by taking into account some fixed and flexible restrictions of cloud computing user's jobs.

Bayati M et.al [8] have introduced two iterative, distributed algorithms for scheduling, Є- Auction algorithm Є-min-sum algorithm. Their simulation results show that both these algorithms are throughput and delay optimal, which makes them suitable for scheduling in networks.

Shu-Ching et.al, [19] have proposed a Three-phases scheduling algorithm in hierarchical cloud computing that integrates Best Task Order, Enhanced Opportunistic Load Balancing and Enhanced Min-Min scheduling. In the first phase the Best Task Order scheduling algorithm is proposed, which determines the execution order for each task request and hence enhances the performance of system. In the second phase, Enhanced Opportunistic Load Balancing algorithm is proposed. It assigns a suitable service manager for allocation of the service node. Finally in the third phase an Enhanced Min-Min scheduling algorithm is proposed, which guarantees that the suitable service node assigned will execute the task in the minimum execution time. The Enhanced Min-Min enhances the performance of the system. The Optimistic Load Balancing and Min-Min enhances the system performance by 50% while combination of Enhanced Optimistic Load Balancing and Min-Min enhance performance by 20%. An overall proposed scheduling recognizes the load balance of nodes and enhances the entire execution performance of the system.

S. Selvarani et.al, [21] have compared their proposed scheduling algorithm with Activity based costing algorithm in cloud. They have presented activity based costing of the resources and the computation performance. In cloud computing, each application will run on a virtual system, where the resources will be distributed virtually. Every application is completely different and is independent and has no link between each other for e.g. some require more CPU time to compute complex task, and some others may need more memory to store data etc. Resources are sacrificed in activities performed on each individual unit of service. In order to measure direct costs of applications They have proved that the Improved Activity based costing algorithm has reduced the processing cost of the tasks submitted to the cloud system. They have measured direct cost of applications and every individual user of resources like CPU cost, memory cost, input/output cost etc. When direct data of each individual resources cost has been measured, more accurate cost and profit analysis based on it than those of the traditional way can be achieved.

Volker Hamscher et. al [6] have provided a broad view about the role of task scheduling in a Grid computing environment. The scheduling structures that occurs in Grid computing with scheduling algorithms and their selection strategy applicable to differing structures leads to centralized and de-centralized schedulers. In a centralized environment all parallel machines are scheduled by a central instance, information on the state of all available systems must be collected from here. Due to the complexity in synchronization and simultaneous execution in decentralized scheduling, centralized scheduling is more preferred over decentralized scheduling, also former scheduling method is able to produce very efficient schedules, because the central instance has all necessary information on available resources and even different policies can be used for local and global job scheduling. The decentralized scheduling provides better fault-tolerance and reliability as failure of single component will not affect the whole metasystem. The scheduler of a metacomputing environment usually arranges the submitted jobs in order to achieve high utilization. The task of scheduler in metacomputing environment is more complex as many machines are involved with mostly local scheduling policies. They have also presented metacomputing scheduler must therefore form a new level of scheduling which is implemented on top of the job schedulers and it is likely that a large metacomputer may be subjected to more frequent changes as individual resources may join or exit the grid at any time.

Maleeha Kiran et. al  [9] have presented an overview of modeling and performance evaluation of hierarchical job scheduling on the Grids. They have found a common problem arising in to select most efficient resource to run a particular program, wherein users are required to reserve in advance the resources needed to run their program submission depends on guesswork by user, which leads to insufficient use of resources, incurring extra operation cost such as idling queues or machines. Thus they have designed a prediction module to aid the user. Optimal allocation of resources to the submitted jobs based on predicted execution time of the program using various aspects of static analysis, analytical benchmarking and compiler based approaches.

Luqun Li [11] have proposed an optimistic differentiated service job scheduling system for cloud computing service users and providers. For various QoS requirement of cloud computing users job, he builds non-preemptive priority M/G/1 queuing model for the jobs, then considering the cloud computing service providers destination which is to gain the maximum profits by offering cloud computing resources, a system cost function for this queuing model is build. Considering the goals of both cloud computing users and service providers based upon his queuing model and system cost function, an optimistic value of service is provided for each job in the corresponding non-preemptive priority M/G/1 queuing model. His approach for job scheduling system in cloud computing environment not only achieve QoS requirement of the cloud computing users job but also can make the maximum profits of the cloud computing service provider.

Rasooli A et. al, [10] have introduced a novel rule based algorithms for scheduling in Grid computing systems. This algorithm maintains information not only about the resources but also about the run time of all jobs in a scenario where jobs arrive over the time and disappear from the scheduling process at their completion time. They have implement several dynamic scheduling algorithms by combination of some dispatching rules and evaluated and compared their efficiency under various criteria such as Makespan, Tardiness, Flow time etc. They have divided the scheduling algorithms into two categories, static scheduling algorithms and dynamic ones. In static mode, every job is assigned once to a resource. Thus placement of application is static and a firm estimate of the cost of the computation can be made in advance of the actual execution. Dynamic scheduling is usually applied when it is difficult to estimate the cost of applications online dynamically. They have presented advantage of dynamic load balancing over static scheduling is that the system need not be aware or the run-time behavior of the application before execution. They have divided the process of scheduling a job into two phases. First phase a proper resource should be selected according to the requirements of the coming job and the properties of the resources. In Second phase, the incoming jobs should be located in an appropriate place of the selected resource's queue. Based upon analysis on these two phases of job scheduling they have introduced a new dispatching rule for resource selection in the grid called MM*. Also they have presented another new dispatching rule for the first stage of scheduling called Minimum Schedule Completion.

Yi Wei and M. Brian Blake [16] proposed combination of service oriented computing and cloud computing. They have presented challenges for this coordinated computing, Maintaining High Service Availability, providing end to end secure solutions, managing longer standing service work – flows, rapid service deployment i.e. cloud computing providers should add features to their cloud infrastructures to enable management and monitoring for deployed services. These management and monitoring functionalities should not only consider the status of deployed services but also take into account the status of underlying cloud infrastructures. Service level agreements in future integrated services and cloud systems will operate with more accuracy and confidence.

## III. FRAME WORK AND METHODOLOGY

Figure 1 presents a Cloud Computing environment. It is a pool of distributed resources which is accessed by a cloud computing users. Generally, the cloud computing users can specify the due time (deadline) for the accomplishment of their submitted tasks. It is the responsibility of the cloud computing service providers to provide the adequate Service Level Satisfaction (μ) so that the available resources are allocated in order to achieve the QoS along with costumer relationship values for the best resources utilization.

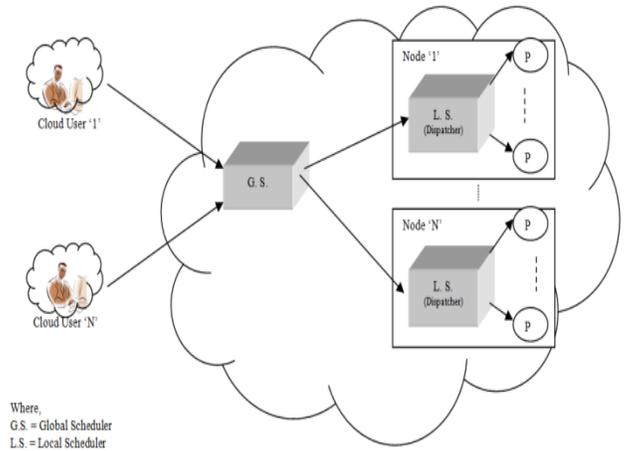

Figure 1. An Illustration of Scheduling at different levels in Cloud Computing Environment.

In Cloud Computing, a set of hypothesis can be arrived at for the quality of delivered services as follows:

Hypothesis
- Jobs shall be accepted and acknowledged within stipulated period of time.
- Job shall be completed and delivered within the requested time.
- Any job completed ahead before deadline is considered as good QoS.
- Any job slipped deadline is considered as poor QoS.

The Service Level Satisfaction (μ) of a task in a queue is the rate by which task is allocated for execution providing that the efficient resources is present in the cloud computing environment. We have considered that queue-1 shall have high service level satisfaction than the subsequent queues. From the cloud computing service provider prospective, we have taken an M/ G/ 1 queuing model to distinguish the incoming task. Here each job queue will have separate Service Level Satisfaction (μ) that shall be calculated based upon the Technical Priority ($T_p$) and the Business Priority ($B_p$), as shown in equation 1

$$\mu i = f(T_p + B_p) \ldots\ldots\ldots\ldots\ldots\ldots\text{Equation 1}$$

Where,

$T_p$ = Technical Priority,
$B_p$ = Business Priority.
i = Number of Job Queues.

The tasks that submitted to the cloud is in accordance with the Poisson distribution with rate λ and the task scheduling system in Cloud assigns the incoming tasks in the form of general distribution [10].

$$\lambda = \sum_{i=1}^{n} \lambda i$$

Where, n is the no. of task arrived.

Suppose if each task has its due time $T_i$ and the $T_{i\text{-tolerant}}$ is the maximum tolerant delay for a cloud computing user task, then to meet the QoS requirement,

$$T_{i\text{-tolerance}} << T_i$$

The tolerant time $T_{i\text{tolerance}}$ is,

$$T_{i\text{-tolerance}} = T_d + T_e + T_{RAP}$$

Where,
$T_d$ = Time involved in Task Delivery
$T_e$ = Task Execution Time
$T_{RAP}$ = Resource allocation and preparation time for task.

Here, for ordering the priority of tasks we are considering few aspects such as the start time and total number of jobs arrived in an immediate past time frame. Equation 2 estimates the start time of the task. The start time of a task is calculated based on Due time ($T_d$), Execution time ($T_e$), Preparation time ($T_{prep}$) and a Blank time ($T_b$).

$$T_{start} = T_d - T_e - T_{prep} - T_b \ldots\ldots\ldots \text{Equation 2}$$

Blank wait time of tasks experienced for resources is a statistical value that changes over a period of time. At zero overload Blank wait time tends to be zero and at full overload or in situation of resources crises, Blank waiting time is defined by a feedback value available from resource allocation manager. We are here assuming the Blank waiting time to be zero i.e. there is no overload at the time of this analysis.

The preliminary priority order follows a 'Tstart' order ascending in realtime hence the job which has the earliest start having highest priority and the job of last start will have the least priority.

At the preliminary priority estimation phase it is not possible to calculate job resource demand but a heuristics estimation is possible. Current heuristics used is based on the feedback from the resource manager statistics as shown below in Table 1[13].

TABLE I.    TYPE OF RESOURCES AVAILABLE IN CLOUD(AMAZON EC2)

| Type of Resource Available (Instance) | Resource Characteristics | | | | |
|---|---|---|---|---|---|
| | Cores (ECUs) | RAM | Arch. (Bits) | Disk (GB) | Cost ($/h) |
| m1.small | 1(1) | 1.7 | 32 | 160 | 0.1 |
| m1.large | 2(4) | 7.5 | 64 | 850 | 0.4 |
| m1.xlarge | 4(8) | 15.0 | 64 | 1690 | 0.8 |
| c1.medium | 2(5) | 1.7 | 32 | 350 | 0.2 |
| c1.xlarge | 8(20) | 7.0 | 64 | 1690 | 0.8 |

Where, m1 and c1 are the instances with different configurations as medium, large and xlarge.

A resource availability driven parameter takes care of the resource trend in cloud which is managed by the cloud resource allocation manger. Resource Demand Trend or the Resource Demand Weight for a number of jobs arrived in a specified time slot is calculated based upon Number of required processor ($P_n$), memory required ($M$) and Storage Space required ($S$).

If $P_{Rjn}$ is the contribution to Priority due to Resources $T_{Jtn}$ be the total jobs arrived for time frame $t_n$

$$T_{Jtn} = P_n + M + S \ldots\ldots\ldots\ldots \text{Equation 3}$$

The parameters in equation 3 are maintained by the Resource Allocater. If the resource demand is higher for an incoming task then higher priority value will be allocated to that job.

The Technical Priority $T_p$ i.e. starting priority of each task is decided based upon technical aspects of job. It is a function of execution demand and the resource demand.

$$\text{Technical Priority } (T_p) = fn \, (T_{start}, \, P_{rjn}) \ldots\ldots \text{Equation 4}$$

The business priority ($B_p$) shall be decided based upon the aggregate financial transactions till date and the good-will of the costumer with current Cloud service provider.

$$B_p = A_0 * C + B_0 * R \ldots\ldots\ldots \text{Equation 6}$$

Where,
C = Total amount of current order.
R = Total amount of relationship factor involved.
$A_0$ and $B_0$ are the normalization factors to control maximum value of priority modulation that we want to achieve.
We have defined the threshold limit ($\beta$) for both technical priority ($T_p$) and business priority ($B_p$) within which the task shall be executed with its native incoming priority.

Whenever the threshold limit is crossed beyond the defined value than the total priority shall be calculated as presented in below table II.

TABLE II.    SERVICE LEVEL SATISFACTION AND RESULTANT PRIORITY CALCULATION BASED UPON BUSINESS PRIORITY

| Types Of Priority | | | |
|---|---|---|---|
| Technical Priority | Resultant Priority | S.L.S(μ) (with Tp) | S.L.S(μ) (with Bp) |
| 80 | 90 | 80 | 90 |
| 78 | 88 | 78 | 88 |
| 76 | 86 | 76 | 86 |
| 74 | 84 | 74 | 84 |
| 72 | 82 | 72 | 82 |
| 70 | 80 | 70 | 80 |
| 68 | 78 | 68 | 78 |
| 68 | 76 | 68 | 76 |
| 66 | 74 | 66 | 74 |
| 62 | 70 | 62 | 70 |
| 60 | 60 | 60 | 60 |
| 58 | 58 | 58 | 58 |
| 56 | 57 | 57 | 57 |
| 54 | 55 | 55 | 55 |
| 52 | 53 | 53 | 53 |

TABLE III.    SIMULATION SETUP

| Simulation Setup | |
|---|---|
| Parameters | Units |
| Number of Tasks | 2000 |
| Number of VMs in Cloud | 2500 |
| Due Time | 700 sec |
| Execution Time | 650 sec |
| Preparation Time | 5 sec |

TABLE IV.    RESOURCE ALLOCATION VS JOB PRIORITY

| Resource Allocation Statistics | |
|---|---|
| Resource Allocation Probability | Job Priority |
| 1 | 1-10 |
| 1 | 11-20 |
| 0.9 | 21-30 |
| 0.9 | 31-40 |
| 0.8 | 41-50 |
| 0.7 | 51-60 |

So, the total Resultant Priority based on demand shall be,

$$T_{resultant} = T_p + B_p$$

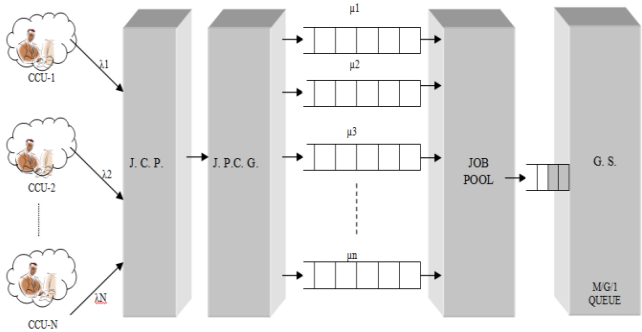

Where,
CCU = Cloud Computing User
'λᵢ' = Poisson Distribution Rate for Incoming Jobs
J.C.P = Job Collection Point
J.P.C.G. = Job Priority Classification Gate
'μ' = Service Rate
G.S. = Global Scheduler

Figure 2.Characterization of incoming Tasks in Prioritizing Unit.

## IV PERFORMANCE ANALYSIS AND RESULTS

### Simulations Setup

We have created our own simulation setup using object oriented programming approach and available data as shown below in Table III. Table IV shows the Resource Allocation Statistics and the related graph is shown in figure 3. The probability of Tasks with highest priority are allocated the requested resources.

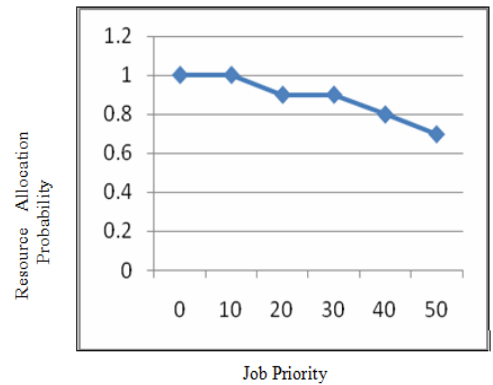

Figure 3. Normal Scenario for resource allocation for Incoming Jobs

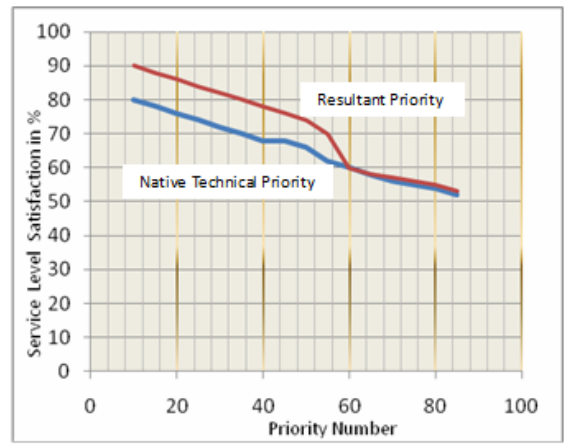

Figure 4 Service Level Satisfaction obtained with Native Technical Priority and Resultant Priority.

Since a single priority chain is insufficient, we consider two dimensional priority chains (m, n), where m and n ranges form ([1, N]) where 1 is the highest priority and subsequent priority values shall be the lower ones. So, in the case task of priority $T_{p(1,5)}$ is much better than $T_{p(2,1)}$. As shown in Figure 2 the incoming task is collected and acknowledged at the job collection point (JCP). Once the job is collected it shall be prioritized based upon the technical priority and business priority. Figure 4 shows the Service Satisfaction Level on Native Priority of tasks and figure 5 shows the Modified Service Satisfaction Level ($\mu$) with adaptive business priority.

With the estimation of the Modified priority, the waiting hours of the jobs submitted to the cloud system is observed. The jobs with higher priority have less waiting time. The graphs in figure 5 and 6 show the waiting time in hours.

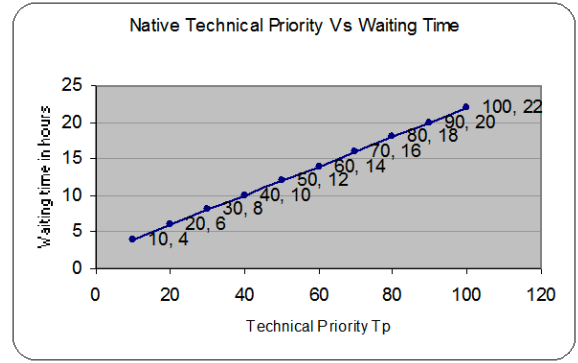

Figure 5 Native Technical Priority Vs Waiting time

TABLE V.      NATIVE TECHNICAL PRIORITY VS WAITING TIME

| Native Priority | Waiting time (in hrs) |
|---|---|
| 10 | 4 |
| 20 | 6 |
| 30 | 8 |
| 40 | 10 |
| 50 | 12 |
| 60 | 14 |
| 70 | 16 |
| 80 | 18 |
| 90 | 20 |
| 100 | 22 |

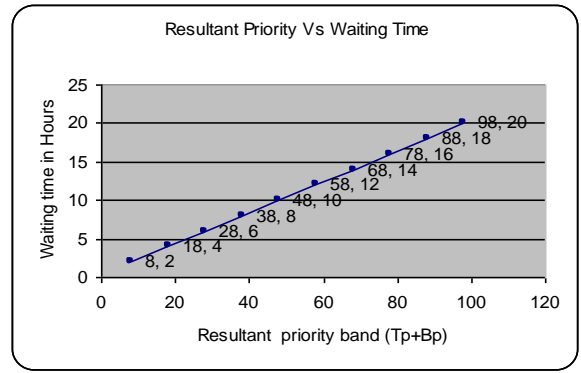

Figure 6 Resultant Priority Vs Waiting time

TABLE VI.      RESULTANT PRIORITY VS WAITINGTIME

| Resultant Priority | Waiting time (in hrs) |
|---|---|
| 8 | 2 |
| 18 | 4 |
| 28 | 6 |
| 38 | 8 |
| 48 | 10 |
| 58 | 12 |
| 68 | 14 |
| 78 | 16 |
| 88 | 18 |
| 98 | 20 |

V.  CONCLUSION AND FUTURE WORK

In this paper we put forward a theoretical approach to finalize the priority of each job that arrives in to the cloud computing environment, taking in to account the business value and current trend of resource availability. The simulation studies have shown that priority is getting modulated with ample importance to the resource availability and business value of each job. The priority produced like this give enough support for a resource manager, scheduler working towards executing jobs on virtual machines. Future work in this line can create estimation methods of business priority value and functions of hybridization of technical priority and business priority for meeting the business needs of clouds of various formats.